# Numerical optimization of single-mode fiber-coupled single-photon sources based on semiconductor quantum dots

LUCAS BREMER,[1] CARLOS JIMENEZ,[2] SIMON THIELE,[2] KSENIA WEBER,[3] TOBIAS HUBER,[5] SVEN RODT,[1] ALOIS HERKOMMER,[2] SVEN BURGER,[4] SVEN HÖFLING,[5] HARALD GIESSEN,[3] AND STEPHAN REITZENSTEIN[1,*]

[1]*Institute of Solid State Physics, Technische Universität Berlin, 10623 Berlin, Germany*
[2]*Institute for Applied Optics (ITO), University of Stuttgart, 70569 Stuttgart, Germany*
[3]*4th Physics Institute, University of Stuttgart, 70569 Stuttgart, Germany*
[4]*Zuse Institute Berlin (ZIB), 14195 Berlin, Germany & JCMwave GmbH, 14050 Berlin, Germany*
[5]*Technische Physik, Physikalisches Institut and Wilhelm Conrad Röntgen-Center for Complex Material Systems, Universität Würzburg, 97074 Würzburg, Germany*
**stephan.reitzenstein@physik.tu-berlin.de*

**Abstract:** We perform extended numerical studies to maximize the overall photon coupling efficiency of fiber-coupled quantum dot single-photon sources emitting in the near-infrared and telecom regime. Using the finite element method, we optimize the photon extraction and fiber-coupling efficiency of quantum dot single-photon sources based on micromesas, microlenses, circular Bragg grating cavities and micropillars. The numerical simulations which consider the entire system consisting of the quantum dot source itself, the coupling lens, and the single-mode fiber yield overall photon coupling efficiencies of up to 83%. Our work provides objectified comparability of different fiber-coupled single-photon sources and proposes optimized geometries for the realization of practical and highly efficient quantum dot single-photon sources.

## 1. Introduction

The dawn of the second quantum revolution asks for practical solutions for the real-world implementation of innovative ideas and concepts that have so far mainly been pursued in lab environments. Among them are deterministic quantum light sources which are key elements of quantum communication networks [1] and photonic quantum computers [2]. In this context, quantum light sources provide photons that act for instance as flying qubits - ideally on-demand. For building long-distance quantum networks, the quantum repeater protocol [3] is most promising, providing the quantum equivalent of classical optical amplifiers by means of entanglement swapping [4]. The protocol requires that the photons propagate in a defined spatial-temporal and spectral mode, which makes the use of single-mode fibers mandatory [5, 6]. Moreover, on-chip fiber-coupled single-photon sources (SPS) offer potentially great advantages in terms of long-term stability for fiber-based quantum communication networks [7].

Different quantum emitters have been considered for realizing deterministic SPSs. Promising candidates include NV [8] and SiV [9] centers in diamond, localized emitters in 2D quantum materials [10], and semiconductor quantum dots (QDs) [11]. An overview of the different systems is given in [12]. While exciting results on QD based SPSs with close-to-ideal emission properties have been achieved in recent years [11] and success is made in shifting the emission wavelength to the telecom O-band [13, 14, 15] and C-band [16, 17], powerful solutions for their efficient fiber-coupling are still pending despite first promising results. This includes fiber-coupled QD sources based on microcavities [18, 19], mesa structures [13], photonic crystal nanobeams [20], and optical waveguides [21, 22]. So far, the coupling efficiency of such solutions, in which the fiber is usually brought into direct contact with the structure is limited mainly by poor mode matching with the optical fiber. This severe issue can be mediated by using far-field coupling with additional optical elements to maximize the mode-matching between the quantum light source and the fiber core. In order to maintain the important "plug-and-play" characteristic in the case of far-field coupling, the normally used bulky free-space optics must be substituted by a small footprint fiber-coupling on-chip implementation. First results in this direction were reported in Ref. [23] in which a QD-microlens was on-chip fiber-coupled using a 3D printed microlens system.

To account for structures with sizes in the wavelength range, Maxwell's equations must be solved rigorously in this process to provide an accurate modeling of the devices. Based on this, single-mode fiber-coupling has been studied in the past, for example, for the telecom O-band for a microlens or mesa [24], for an electrically-driven micropillar [25] and for a circular Bragg grating (CBG) [26]. However, these works did not include additional optical elements to maximize the source-fiber mode-overlap which is necessary to reach the optimum performance in terms of fiber-coupling efficiency. Furthermore, it should be noted that in the case of near-field coupling, which is not encountered here, the coupling efficiency is easily overestimated because the complex near-field pattern results from a superposition of modes and a calculation of the field overlap with the fundamental modes of the fiber often neglects that not all of the superpositioned modes are inevitably carried by the fiber [25]. Far-field coupling was recently considered for NV centers in diamond, however, using a black-box optical imaging system for coupling the emitted far-field to the optical fiber [27].

In this work, we report on comprehensive numerical studies on maximizing the photon extraction efficiency ($\eta_{\mathrm{ext}}$) and fiber-coupling efficiency of QD-based SPS. For all modeled structures and emission windows, the mature and for SPSs most advanced GaAs material system is used [11], with embedded InAs/In(Ga)As QDs as quantum emitters. The simulations are performed using the finite element method (FEM) and consider the micromesa, microlens, CBG, and micropillar concepts combined with an intermediate lens for maximizing the source-fiber mode overlap. The four different concepts are benchmarked against the overall photon-in-fiber-coupling efficiency $\eta_{\mathrm{total}}$ while taking also practical and technical aspects in their fabrication into account. $\eta_{\mathrm{total}}$ is defined as the fraction of the generated photons per excitation event which are extracted from the structure and are coupled into the fiber. An internal quantum efficiency of 1 is assumed for the QD emitters [28].

Even though all coupling systems presented are specific to one fiber type, they can be adapted to any fiber type, making our method likewise unique and compatible with any existing fiber network. Our quantum device design optimization is carried out in direct coordination with the capabilities provided by established 3D two-photon laser lithography [29, 30, 31]. This ensures that the selected lenses can be practically implemented with high yield [32, 23]. It is noteworthy that the overall system (semiconductor structure – coupling system – single-mode fiber) is computed self-contained.

## 2. Methods

### *2.1 Parameterization of the numerical models*

In the following, the mentioned four prominent concepts of SPSs based on semiconductor QDs are considered (see Fig. 1) and optimized with respect to $\eta_{\mathrm{ext}}$ at wavelengths of 930 nm (most common for InGaAs QDs), 1310 nm (telecom O-band), and 1550 nm (telecom C-band) to allow for a fair comparison, especially with regards to the single-mode fiber-coupling discussed in section 3. The systems under study can be roughly divided into two groups that follow different approaches, whereby the transitions can be fluid. On the one hand, a geometric approach is taken in which the semiconductor surface is modulated in such a way that total reflection at the semiconductor-air interface is reduced, thus increasing $\eta_{ext}$. The typical representative for this approach is the monolithic QD-microlens [33, 34]. In the second approach, light-matter interaction effects, described by cavity quantum electrodynamics (cQED), are exploited to increase the brightness of the source. A typical representative is the micropillar, where the QDs are localized in a λ-cavity sandwiched between two distributed Bragg reflectors (DBRs), leading to a vertical confinement of the light field while the refractive index contrast at the semiconductor-air interface at the micropillar sidewall ensures lateral mode confinement [35, 36, 37].

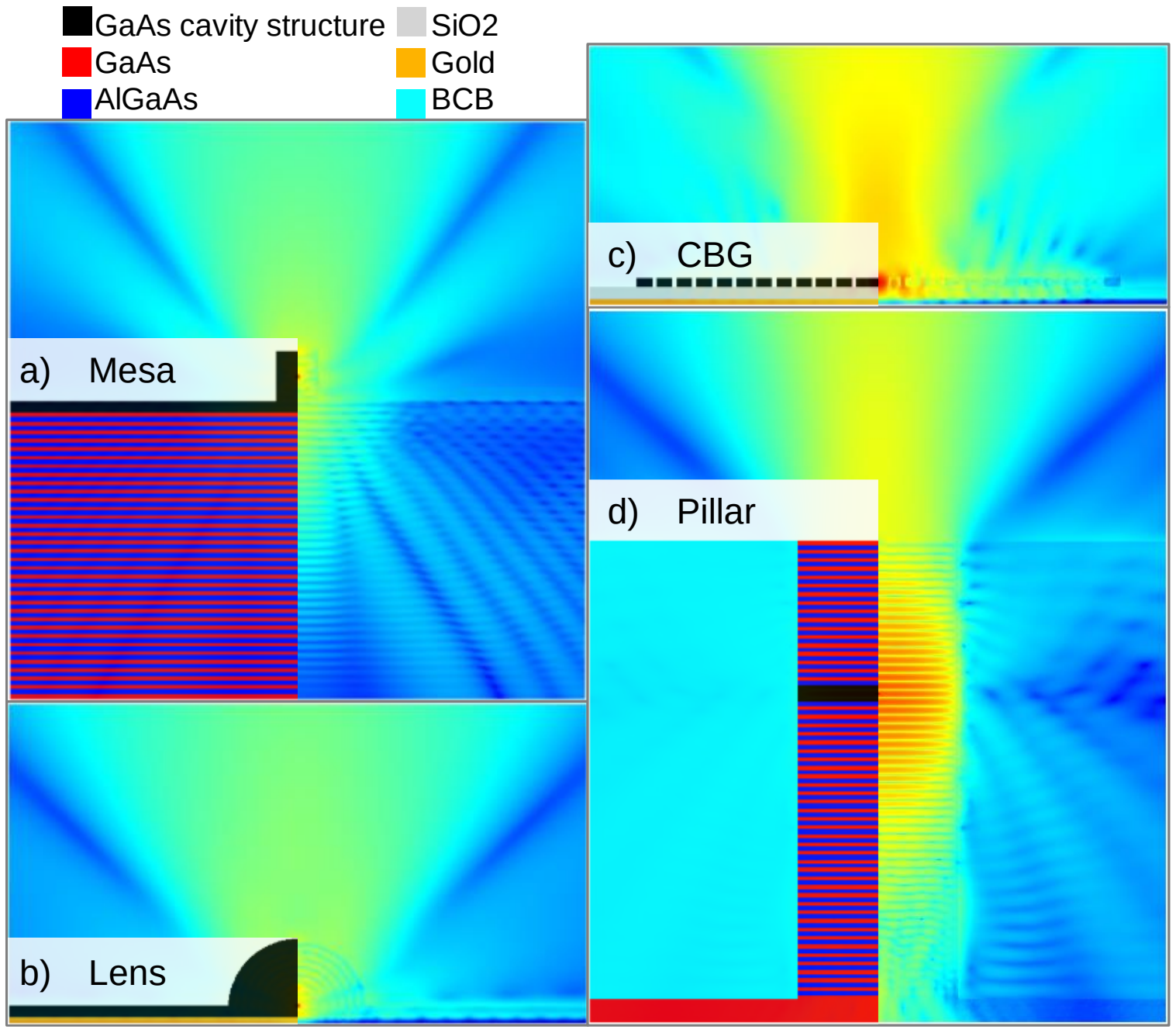

Fig. 1. (a-d) show the layout (cross section) of micromesa, microlens, circular Bragg grating and micropillar structures, respectively, on the left side. From the highest point of each structure, the mapped computational domain shows 4 µm of air. The exemplary structures shown emit at 930 nm, although the geometries are also representative of the telecom wavelengths. The size of the structures scales with increasing wavelength (see tables 1-3). The structures are optimized for maximum photon extraction into a solid angle of 23.6° (NA = 0.4) while providing the highest possible Purcell enhancement. The corresponding electro-magnetic field-intensity distributions calculated with FEM are shown in the background of each sub-figure.

If the QD couples efficiently to the localized micropillar mode, an increase of the spontaneous emission rate is observed due to the Purcell effect [38], enhancing the brightness of the source via a high β-factor and directional emission via a Gaussian far-field pattern. In recent years, SPSs based on a CBG, which spans concentrically around a central mesa, have become very popular and show excellent emission properties [39, 40, 15]. The photon extraction from the central disk is thereby enhanced by the lateral Bragg grating. In addition, the very simple micro/nanomesa design [33, 41] has been studied, creating a weak confinement of the lateral light field associated with a moderate Purcell-effect but also surprisingly high photon extraction efficiencies of up to 82% for a numerical aperture (NA) of 1.0 [42, 41]. The mentioned photon extraction strategies differ not only in their radiation properties, but also in spectral bandwidth and manufacturing complexity. Therefore, the choice of the most suitable device concept depends on the respective application. For instance, broadband CBGs are particularly interesting for the generation of polarization-entangled photon pairs [39, 40], while geometric concepts may be advantageous, e.g., for coupling with atomic ensembles, where a pronounced Purcell factor ($F_P$) and thus a broadening of the homogeneous linewidth would be counterproductive [43].

The epitaxial layer structure of the mesa is the same for 930 nm and 1310 nm, so that only the 930 nm case is shown on the left side of Fig. 1a. The layer sequence of the structures at 1550 nm differs slightly from their shorter wavelength counterparts, since a metamorphic buffer layer is needed to shift the emission wavelength of the InAs QDs into the telecom C-band [44]. It has recently been shown that it is possible to reduce the initially applied thickness of the buffer layer from over 1 µm to ≥ 180 nm [45], which allows the formation of equivalent structures also for the telecom C-band. Further details are given in the SI. In all cases, the micromesa is located above a distributed Bragg reflector (DBR) consisting of 25 AlAs/GaAs λ/4-thick layer pairs. The following length scales were considered for the optimization of the photon-extraction efficiency: Mesa radius and height, GaAs spacer layer between mesa and DBR and vertical position of the dipole point source relative to the mesa-spacer interface representing the quantum dot. For the microlens and the CBG structure, a gold layer was used as bottom mirror, as has been reported in previous experimental implementations [33, 39, 40, 41]. The necessary flip-chip process is more demanding than the growth of a DBR but offers the advantage of (more or less) wavelength independent high reflectivity even for larger angles of incidence. Moreover, it facilitates strain tuning of the QD emission properties when combined with piezo-actuators via thermo-compression Au bonding [46, 47] and can serve as a backside electrical contact.

The microlens (see Fig. 1b, left) was defined in the two-dimensional (2D) plane by a curved section with the conic constant and by the radius of curvature and the diameter of the lens. In addition to the three parameters, the spacer thickness and the position of the dipole point source relative to the lens-spacer interface were also optimized. However, the last two parameters were not optimized simultaneously because of the strong mutual interdependence and the wide range of conceivable lens geometries which opens up a parameter space that can hardly be handled properly due to its large number of local maxima of $\eta_{ext}$. The CBG and micropillar structures rely on cQED effects, which can be described analytically in good approximation. Therefore, the position of the dipole was not varied here, since it was assumed that the best performance results from a central position in the cavity, i.e., at the antinode of the confined light field. For the CBG structure (Fig. 1c, left), the height of the central disk and rings, the diameter of the central disk, the grating period, the trench width between the rings, and the thickness of the low-refractive-index $SiO_2$ layer sandwiched between the semiconductor membrane and the gold mirror were optimized for maximum $\eta_{ext}$ into collecting optics with NA = 0.4. For the micropillar (Fig. 1d, left), only the micropillar radius and cavity height were optimized. The studied microcavity is formed by 35 (17) AlAs/GaAs λ/4-thick layer pairs in the lower (upper) DBR to maximize $\eta_{ext}$. The micropillar is planarized with benzocyclobutene (BCB) polymer, as is often used for mechanical stabilization and sealing of the sidewall.

### *2.2 Efficient optimization through self-learning algorithm*

For the numerical simulations, JCMsuite was used, which provides a commercial solver for the Maxwell equations based on FEM in the frequency domain [48, 49]. Since the single-mode fiber-coupling is strongly intensity and phase dependent, a high polynomial degree between 3 and 5 was chosen for the ansatz functions. This is important for the calculation of the Purcell effect in the cavity structures because the reliable computation of the Purcell factor requires sufficiently high numerical resolution of the FEM problem. Further, the computational domain may include the source, the intermediate mode-matching lens, and the incoupling facet of the single-mode fiber. It is therefore relatively large, containing areas with very small features as well as extended areas of wave propagation with > 100 µm extension. FEM allows to employ an unstructured mesh which contains small mesh elements in regions of small structures, and which may contain larger mesh elements in homogeneous regions. JCMsuite allows to use higher polynomial degrees of the finite-element ansatz functions for larger triangles than for small triangles (so-called hp-methods [50]). This results in an accurate treatment of wave propagation simulations on such relatively large computational domains.

The QD is represented in the simulations by a point-like, time-harmonic electric current density (dipole-like) lying in the horizontal plane. This is feasible because the dimensions of a semiconductor QD are small compared to the wavelength. To reduce computation time, the rotational symmetry of the geometries is exploited, and the full 3D solution to Maxwell's equation is obtained as a sum of rotationally symmetric fields [24]. In order to account for the real application conditions of semiconductor QD light sources, the refractive indices were estimated for cryogenic temperatures (see SI). After the electromagnetic tensor fields in the computational domain have been calculated, the analysis is performed by means of various post-processing operations. These include in particular the export of field distributions in the form of 2D cross sections, the calculation of the total radiated power of the dipole, the evaluation of the power flux in the far-field as a function of the radiation angle and the calculation of the overlap integral of the radiated field with the guided modes of the fiber. The Purcell factor is calculated as the ratio between the total radiated dipole power and the emission of a dipole in the homogeneous matrix material of the QD (GaAs) without a photonic structure.

The high-dimensional parameter space of the source optimization makes fixed-step parameter scans impractical due to finite computational resources and associated long simulation times. Therefore, a Bayesian algorithm based on Gaussian processes was used for the numerical optimizations [51]. The algorithm, belonging to a class of machine learning algorithms, maximizes (minimizes) an unknown objective function. The objective function is based on a stochastic model, which decides based on the previously calculated evaluations at which point in the high-dimensional parameter space the function should be evaluated next. In the present case, the extraction efficiency into an NA of 0.4 and the Purcell factor were maximized simultaneously, with different weights given over the large number of optimization runs performed. This is illustrated by Fig. 2, which shows the optimization run for the micropillar system emitting at 930 nm. The micropillar system was chosen for the sake of clarity because the parameter space consists of only two parameters (cavity height and micropillar radius) and the progress can thus be displayed. In the case of the micropillar, the function $f = \frac{\eta_{\mathrm{ext}}^2}{85} + \frac{F_{\mathrm{P}}}{200}$ was maximized. The function was chosen based on empirical observations of the correlation of $\eta_{\mathrm{ext}}$ and $F_{\mathrm{P}}$ for the micropillar and is deliberately strongly dominated by $\eta_{\mathrm{ext}}$. In (a) one sees that already with the 45th iteration a parameter is found, which features $\eta_{\mathrm{ext}}$ of 91.3% with a simultaneous Purcell enhancement by a factor of 11 and thus comes very close to the later maximum of the optimization function. Due to the random choice of the first evaluation points, it could have been possible that such a value is found later (or earlier), but the performance of the self-learning algorithm becomes clear when looking at the picture after the 200th iteration (b). Although the system was not given any information about the physical regularities, the optimization has moved along the mode of high $\eta_{\mathrm{ext}}$ and completely evaluated the crucial region of the parameter space, as can be seen in the comparison with (c) which shows 8000 iterations. It is also clear from (c) that the choice of the optimization function is often difficult and ambiguous. So, in the range of micropillar radii below 550 nm, four points are evaluated that show Purcell factors above 50. The $\eta_{\mathrm{ext}}$ amounts to a maximum of 39%. The trade-off between $\eta_{\mathrm{ext}}$ and $F_{\mathrm{P}}$ is structure dependent. In particular, for the micromesa and microlens, it proved to be useful to optimize foremost $\eta_{\mathrm{ext}}$ and to set a minimum allowed value for $F_{\mathrm{P}}$, since these broadband structures rely mainly on the geometric enhancement of emission.

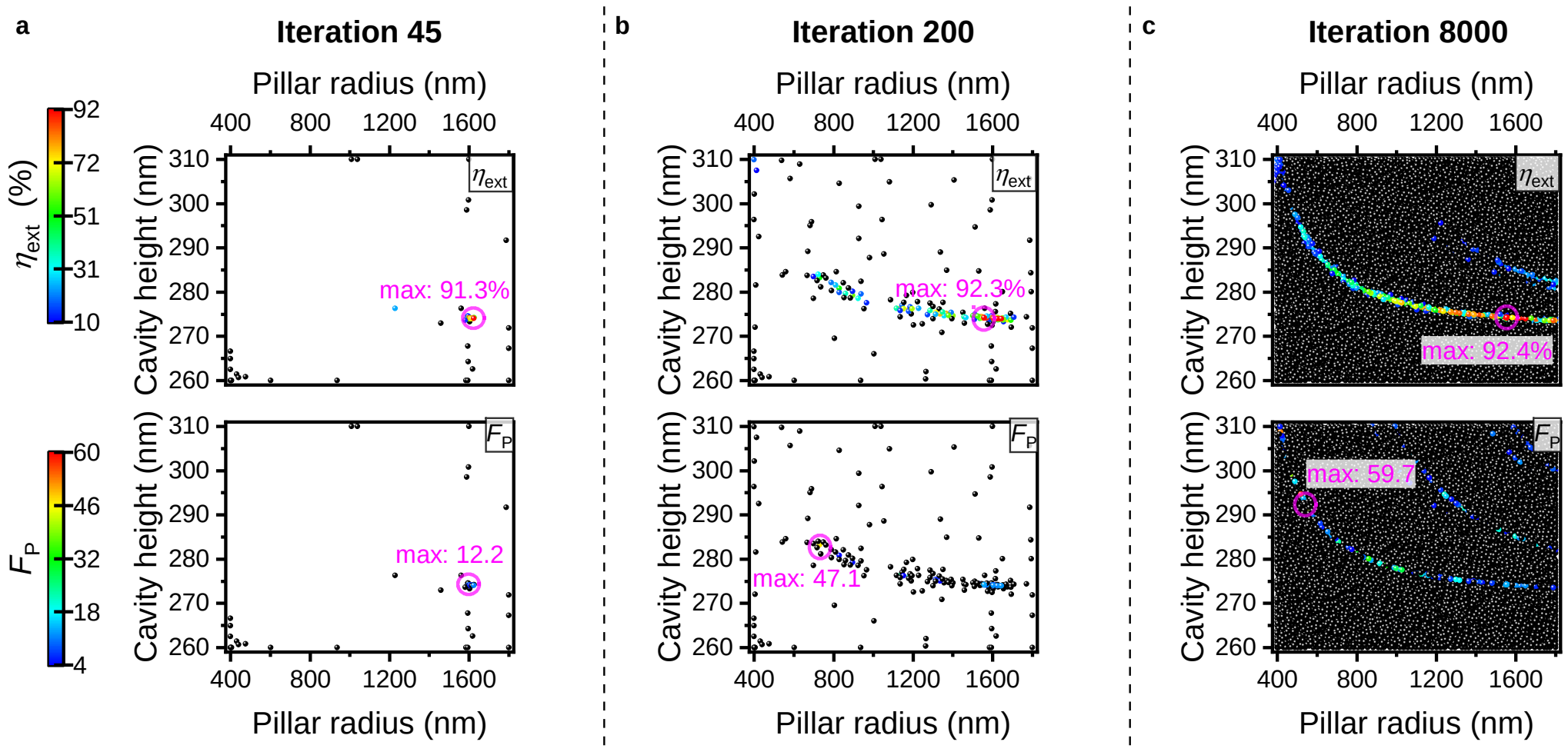

Fig. 2 Optimization progress of the micropillar structure for the emission wavelength of 930 nm. The $\eta_{ext}$ into NA of 0.4 (top) and $F_P$ (bottom) are shown for the 45th (a), 200th (b), and 8000th(c) iteration. Already with the 45th iteration, the algorithm finds a parameter set of cavity height and micropillar radius that yields an $\eta_{ext}$ of more than 90%. Subsequently, the optimization runs along the most efficient mode. After the 8000th iteration, the entire parameter space is sampled, although the sampling density remains significantly increased in areas of high performance. The respective maximum is marked by a magenta-colored circle.

## 3. Results and discussion

### *3.1 Optimized semiconductor structure designs*

The optimized SPS structure design sizes are summarized in Table 1-3 for the different wavelengths. As expected, the structure sizes scale with the emission wavelength, i.e., the longer the wavelength the larger the optimized structure. However, the scaling is not linear due to the dispersion of the refractive indices. In the optimized mesas, it is interesting that the dipole is always placed in the mesa center (in vertical direction). In fact, it should be said that efficient structures can also be found for other dipole positions. For the microlenses, it is striking that the optimal properties were obtained for hemispherical lenses (k = 0), while superhemispherical (Weierstrass) lens shapes were not simulated because they are not compatible with the typical lithography and etching methods used for fabrication. Here, the dipole is located directly at the base plane of the lens. It should be noted that for the 1.3 µm microlens structure, for example, comparable performance (see next section) is obtained by placing the dipole 60 nm below the lens and increasing the GaAs spacer thickness accordingly. It is further striking that the optimal thickness of the GaAs spacer is not structure but only wavelength dependent, although the vertical position of the dipole (as described before) differs fundamentally between micromesa and microlens. Regarding the CBG structures, it should be specified that they have been optimized for air-filled trenches. If the structures are used in liquid helium, a slight red shift of the cavity center wavelength by, e.g., 0.5 nm can be observed for the 930 nm structure, while the performance remains the same.

**Table 1. Optimized structure parameters and optical properties for λ = 930 nm**

| *Structure* | *Optimized parameters (nm)* | *Far-field emission half-angle[2] (°)* | *Purcell factor / Bandwidth FWHM (nm)* | *Extraction efficiency into NA of 0.4 and 0.8 / Bandwidth FWHM (nm)* |
|---|---|---|---|---|
| *Micromesa* | Radius: 370.9<br>Height: 858.5 | (φ = 0°): 26.8<br>(φ = 90°)[3]: 27.7<br>average: 27.3 | $F_P$: 1.1<br>FWHM: broadband[4] | NA 0.4: 55.1%<br>NA 0.8: 79.3%<br>FWHM: broadband[4] |
| | GaAs spacer: 207.1<br>Dipole position[1]: 429.3 | | | |
| *Microlens* | k: 0<br>Radius of curvature: 1214<br>Diameter: 2424<br>GaAs spacer: 202.0<br>Dipole position[1]: 0.0 | (φ = 0°): 23.8<br>(φ = 90°): 16.5<br>average: 20.2 | $F_P$: 1.4<br>FWHM: broadband[4] | NA 0.4: 75.2%<br>NA 0.8: 89.0%<br>FWHM: broadband[4] |
| *CBG* | Height: 157.8<br>Radius central disk: 388.8<br>Grating period: 346.0<br>Trench width: 75.0<br>$SiO_2$ thickness: 200.4 | (φ = 0°): 19.6<br>(φ = 90°): 14.8<br>average: 17.2 | $F_P$: 41.4<br>FWHM: 2.3 | NA 0.4: 90.2%<br>NA 0.8: 97.3%<br>FWHM: broadband[4] |
| *Micropillar* | Radius: 1398.0<br>Cavity height: 274.9 | (φ = 0°): 20.6<br>(φ = 90°): 19.2<br>average: 19.9 | $F_P$: 16.5<br>FWHM: 0.1 | NA 0.4: 90.6%<br>NA 0.8: 94.0%<br>FWHM: 0.5 |

**Table 2. Optimized structure parameters and optical properties for λ = 1310 nm**

| *Structure* | *Optimized parameters (nm)* | *Far-field emission half-angle[2] (°)* | *Purcell factor / Bandwidth FWHM (nm)* | *Extraction efficiency into NA of 0.4 and 0.8 / Bandwidth FWHM (nm)* |
|---|---|---|---|---|
| *Micromesa* | Radius: 538.5<br>Height: 1252.0<br>GaAs spacer: 299.9<br>Dipole position[1]: 626.0 | (φ = 0°): 26.7<br>(φ = 90°)[3]: 27.4<br>average: 27.1 | $F_P$: 1.0<br>FWHM: broadband[4] | NA 0.4: 54.8%<br>NA 0.8: 78.6%<br>FWHM: broadband[4] |
| *Microlens* | k: 0<br>Radius of curvature: 1764.0<br>Diameter: 3522.0<br>GaAs spacer: 303.4<br>Dipole position[1]: -1.9 | (φ = 0°): 21.4<br>(φ = 90°): 17.2<br>average: 19.3 | $F_P$: 1.5<br>FWHM: broadband[4] | NA 0.4: 77.4%<br>NA 0.8: 88.4%<br>FWHM: broadband[4] |
| *CBG* | Height: 266.6<br>Radius central disk: 550.8<br>Grating period: 481.9<br>Trench width: 89.8 | (φ = 0°): 18.5<br>(φ = 90°): 13.8<br>average: 16.2 | $F_P$: 71.9<br>FWHM: 1.7 | NA 0.4: 90.4%<br>NA 0.8: 95.7%<br>FWHM: broadband[4] |
| | $SiO_2$ thickness: 245.0 | | | |
| *Micropillar* | Radius: 1980.0<br>Cavity height: 400.4 | (φ = 0°): 20.6<br>(φ = 90°): 19.7<br>average: 20.2 | $F_P$: 10.8<br>FWHM: 0.2 | NA 0.4: 89.0%<br>NA 0.8: 92.5%<br>FWHM: 0.9 |

**Table 3. Optimized structure parameters and optical properties for λ = 1550 nm**

| *Structure* | *Optimized parameters (nm)* | *Far-field emission half-angle[2] (°)* | *Purcell factor / Bandwidth FWHM (nm)* | *Extraction efficiency into NA of 0.4 and 0.8 / Bandwidth FWHM (nm)* |
|---|---|---|---|---|
| *Micromesa* | Radius: 626.1<br>Height: 1456.1<br>GaAs spacer: 359.0<br>Dipole position[1]: 728.1 | (φ = 0°): 26.6<br>(φ = 90°)[3]: 27.3<br>average: 27.0 | $F_P$: 1.0<br>FWHM: broadband[4] | NA 0.4: 54.8%<br>NA 0.8: 78.5%<br>FWHM: broadband[4] |
| *Microlens* | k: 0<br>Radius of curvature: 2056.0<br>Diameter: 4106<br>GaAs spacer: 350.9<br>Dipole position[1]: 0.0 | (φ = 0°): 20.8<br>(φ = 90°): 16.7<br>average: 18.8 | $F_P$: 1.4<br>FWHM: broadband[4] | NA 0.4: 78.9%<br>NA 0.8: 88.2%<br>FWHM: broadband[4] |
| *CBG* | Height: 392.2<br>Radius central disk: 611.3<br>Grating period: 551.1<br>Trench width: 156.3<br>$SiO_2$ thickness: 278.6 | (φ = 0°): 17.8<br>(φ = 90°): 14.5<br>average: 16.2 | $F_P$: 43.3<br>FWHM: 3.3 | NA 0.4: 91.2%<br>NA 0.8: 96.3%<br>FWHM: broadband[4] |
| *Micropillar* | Radius: 2354.6<br>Cavity height: 466.4 | (φ = 0°): 20.5<br>(φ = 90°): 19.5<br>average: 20.0 | $F_P$: 9.7<br>FWHM: 0.3 | NA 0.4: 88.9%<br>NA 0.8: 92.2%<br>FWHM: 1.1 |

[1] relative to the mesa-spacer interface
[2] intensity decrease to $1/e^2$ for azimuthal angles $\varphi = 0°$ and $\varphi = 90°$ of the far-field
[3] strong deviation from gaussian radiation pattern (see SI)
[4] > 15 nm

## 3.2 Optimized emission characteristics

Next, we discuss the optical results of our numerical source optimization. In Fig. 1, the calculated intensity distributions are shown for each considered structure (exemplary for 930 nm emission wavelength) in a logarithmic presentation. Starting from these distributions, the far-field was calculated, and the radiation half-angles (decrease of the field intensity to $1/e^2$) were extracted. The mesa design always has the largest radiation angle of about 27°, regardless of the wavelength. Also, the radiation profile of the mesa shows the clearest deviations from a Gaussian profile, making the radiation angle not well-defined. A Gaussian radiation profile is important for a high overlap with the fiber modes. Line scans through the respective far-field distribution are shown for all calculated structures in the SI. The lens and the micropillar show similar radiation half-angles of around 20°. However, the far-field of the micropillar is closest to the ideal Gaussian profile of all structures examined. The CBG structures radiate into the smallest solid angle. Here the angle is about 16° to 17°.

Next, $\eta_{ext}$ was investigated from the comparison of the radiated power into a solid angle of the far-field relative to the total dipole power. Fig. 3 shows $\eta_{ext}$ for the different SPS designs and wavelengths. The differences between equivalent structures of different wavelengths are again very small. It can be seen that the cavity-based structures show higher extraction efficiencies than the microlens and mesa for small NA, while the relative differences are significantly smaller when the entire upper hemisphere is considered (NA = 1). The cavity structures thus exhibit modes that radiate more weakly to the lateral side. For the classification of the calculated maximum efficiencies on the basis of experimentally achieved values, the NA of the collecting lens must always be considered. The references listed below refer to In(Ga)As-based QD sources emitting in the visible or near-infrared wavelength range and are consequently compared with Fig. 3(a), since QD emitters in the telecom band are not yet technologically mature at a comparable level. Thus, for an NA of 0.4, 18% was measured for a micromesa [33], 29% for a microlens [52], and 79% for a micropillar [35], compared to 55%, 75%, and 91%, respectively, predicted from the simulations. The strong deviations for the micromesa and microlens can mainly be attributed to the fact that these structures benefit strongly from the performed optimization, since the structures used in the Refs. theoretically show $\eta_{ext}$ below 30% (NA = 0.4) [33, 53]. For a CBG structure, the highest value published to date was 85% for an NA of 0.65 [40], which is just below the simulated $\eta_{ext}$ of 97%.

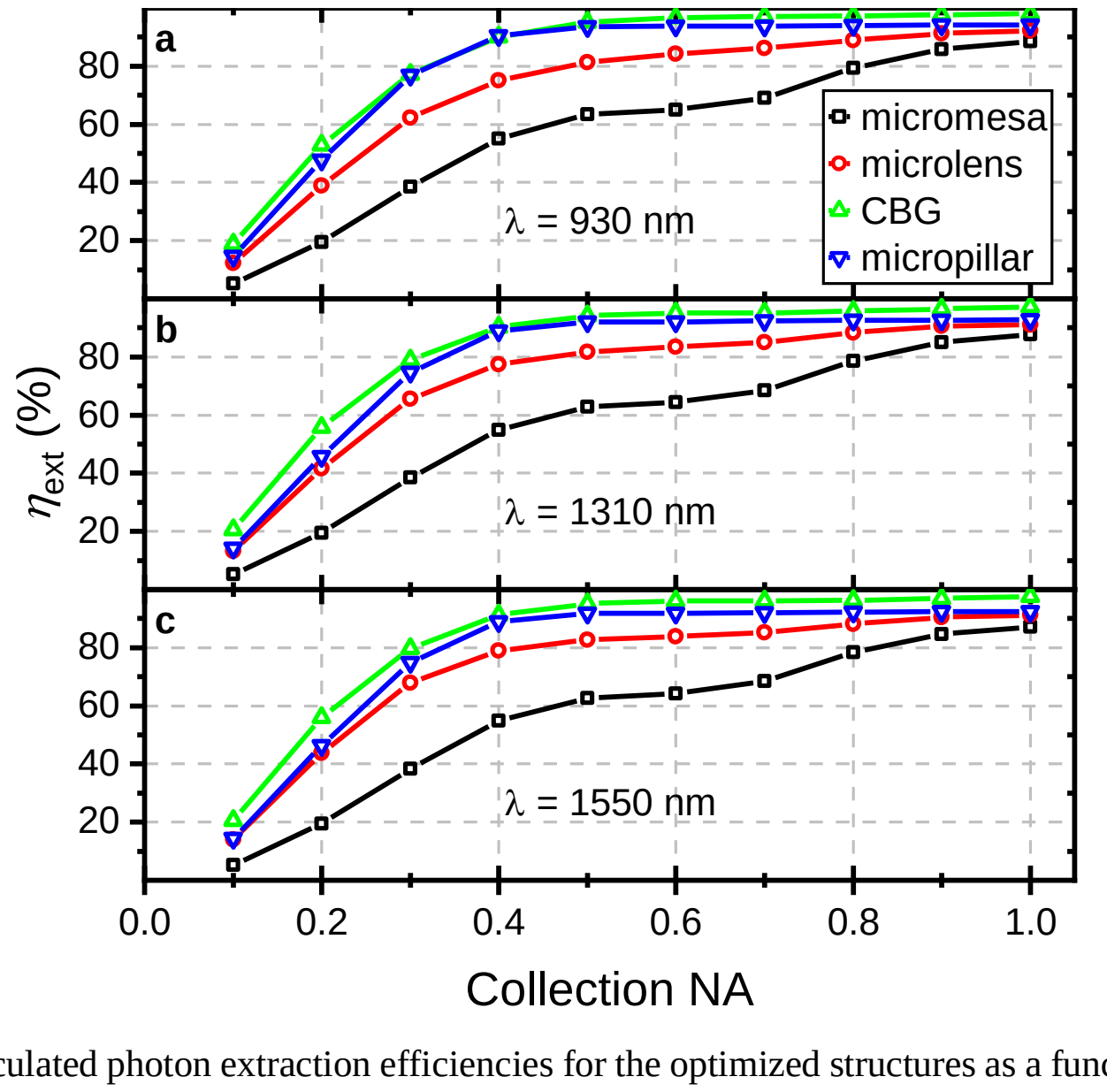

Fig. 3. Calculated photon extraction efficiencies for the optimized structures as a function of the collecting NA for QD emission at 930 nm (a), 1310 nm (b) and 1550 nm (c).

For applications of QD SPSs, a broadband enhancement of emission is desired, such as for the generation of polarization-entangled photon pairs from a biexciton-exciton cascade of a QD [39, 40]. In order to assess this aspect and to ensure comparability of the investigated structures, both $\eta_{ext}$ and $F_P$ were quantified for the different systems as a function of variations of the emission wavelength. The results are summarized in Tables 1-3 in the form of a full width at half maximum (FWHM) of the Purcell enhancement and the extraction spectrum. The underlying spectra can also be taken approximately from Fig. 6, where the total systems including fibers are considered. In addition, Tables 1-3 list the Purcell factors and extraction efficiencies at 930 nm, 1310 nm, and 1550 nm, respectively. The mesas and the lenses always show a low $F_P$ between 1.0 and 1.5. However, for these geometries it is also possible to find structures with significantly higher $F_P$ at the expense of $\eta_{ext}$. For the micropillar $F_P$ is much more pronounced with values between 9.7 and 16.5. Very high $F_P$ could be achieved for the optimized CBGs ($F_P$ = 41.4 to 71.9). It should be noted that the value of $F_P$ = 41.4 for 930 nm could be further increased without changing the extraction efficiency if the trench width were reduced (with adjusted grating period). However, this was not done because the fabrication of such a structure would be difficult to implement with current methods. The bandwidth of the micromesa and microlens is broadband (>30 nm) in terms of both (low) $F_P$ and (high) $\eta_{ext}$. The broadband high $\eta_{ext}$ can also be observed for the CBG, while the Purcell enhancement has a FWHM of a few nanometers. The micropillar system operates in a very narrow band. The FWHM for Purcell enhancement and $\eta_{ext}$ is around 150 µeV (0.1 nm – 0.3 nm) and between 570-720 µeV (0.5 nm – 1.1 nm), respectively.

### *3.3. Optimized overall efficiencies of the fiber-coupled systems*

In this section we focus on maximizing and comparing $\eta_{total}$ of the fiber-coupled QD quantum light sources. To couple the emitted light efficiently into an on-chip fixed single-mode fiber and to avoid interference due to reflections at surfaces with different refractive indices, only a single aspherical lens is considered. Such a lens can be printed with high accuracy directly onto the facet of an optical fiber using 3D direct laser writing [32, 23]. The advantage of a single-lens system, besides the reduced number of writing steps, is that it is not necessary to print it directly on the semiconductor structure to be coupled to the fiber. Hence, we assume that the properties of the quantum light sources, e.g., in terms of the emission wavelength, are not changed by the fiber-coupling, which is necessary in terms of deterministic manufacturing. Furthermore, as few material interfaces as possible are usually preferable in the optical beam path, since this avoids further scattering and reflections.

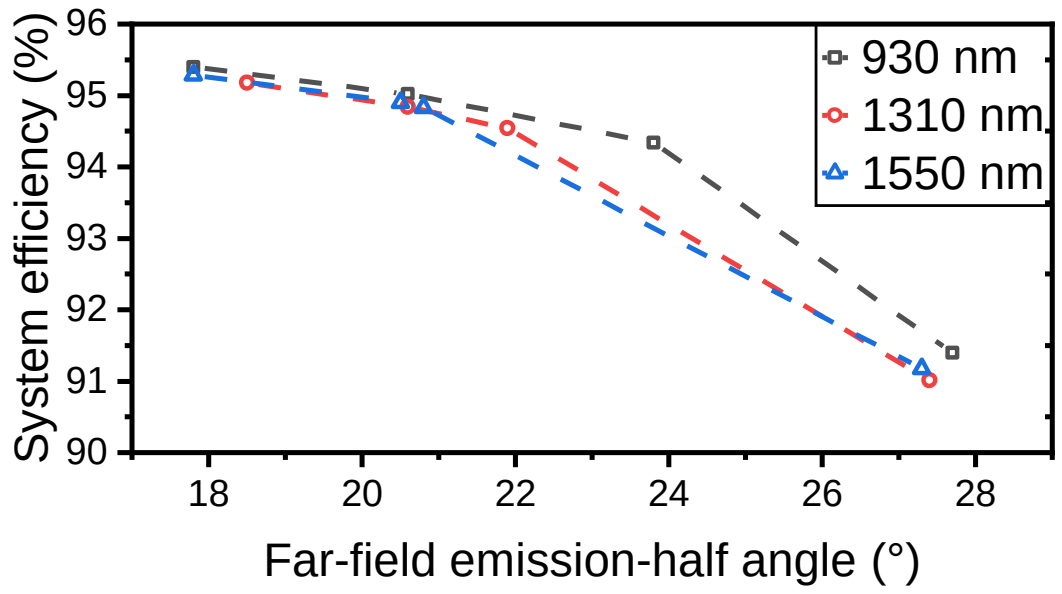

Fig. 4. Maximum achievable system efficiency for single aspheric lens structures as a function of the far-field half-angle of emission. Assumed is a perfect Gaussian source.

The aspherical mode-matching lenses are optimized for each of the four source concepts and three wavelengths and are described by a basic set of functions weighted by aspherical order coefficients. The aspheric lenses were generated by using the optimization routines included in OpticStudio (Zemax) that require the definition of a merit function. For this case, the merit function was based on the estimated total fiber-coupling efficiency which can be described as the product of the system efficiency and the fiber-coupling receiver efficiency. System efficiency accounts for the reflection, absorption, and obstruction losses through the trajectory of the optical field from the source to the intended single mode fiber position. On the other hand, the fiber-coupling receiver efficiency accounts for the overlap between the defined fiber's single mode complex amplitude and the field's complex amplitude generated by the aspheric lens at the position of the single mode fiber. In order to maximize the total fiber-coupling efficiency $\eta_{total}$, it is therefore necessary to maximize these two individual terms. This was achieved by including different system parameters in the optimization routine, such as the aspheric coefficients of the refractive surface, the distance between source field and the apex of the aspherical lens, and others (see SI). It needs to be mentioned that for these single aspheric surface lenses, the far-field half angle of emission imposes a limit on the maximum achievable system efficiency. This dependence can be observed in Fig. 4. Larger emission half-angle values require a stronger degree of angular correction in order to guarantee proper mode matching between the propagating field through the lens and the mode at the receiving fiber side. For this specific case in where single refractive surfaces are used, maximum modal matching can be obtained by using aspheric surfaces with smaller radius of curvature values. Nevertheless, this causes the system efficiency to be generally reduced due to the stronger reflections towards the edges of the aspheric surface. Additionally, the source for each design was parametrized by the free space wavelength and the far-field half-angle of emission, which is obtained from the FEM simulations discussed in Section 2.3. This approach considers the source to be represented by an ideal Gaussian distribution located at the GaAs to air interface. Due to the different beam angles, an individual lens was designed for each source system. The exact lens parameters can be found in the SI.

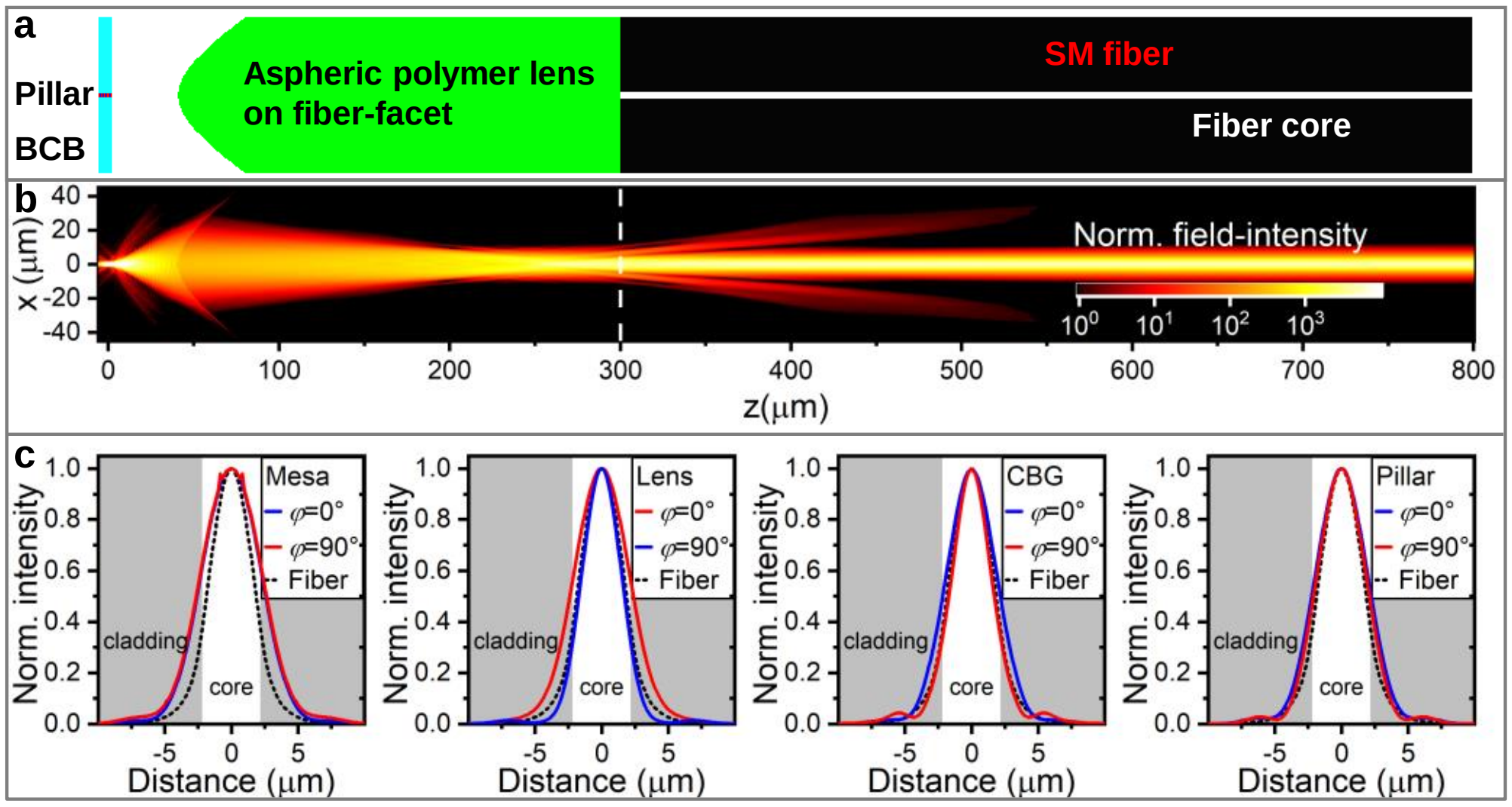

Fig. 5. (a) Computational domain of the entire system including the QD-micropillar emitting at 930 nm, an aspheric polymer lens, and the SM fiber with 500 µm fiber length. (b) Calculated intensity distribution for the full structure shown in a) in logarithmic representation. The aspherical lens focuses the emitted light onto the fiber core, where it is guided in the two fundamental modes. (c) Cross sections through the intensity profiles of the studied structures at 930 nm on the fiber facet for φ = 0° (blue) and φ = 90° (red) (see SI). Dashed lines show the near-Gaussian intensity profile of the fundamental mode of the fiber.

Subsequently, the single-lens systems and the corresponding single-mode fiber were implemented in the FEM solver JCMsuite (see Fig. 5a), and the distance of the coupling system from the photonic structures was optimized by fixed-step parameter scans (see SI). Two of the most common fiber types in the quantum-optical community were used, and the selection was based only on common use and not on possible mode matching. The 780-HP fiber was assumed for the 930 nm and the SMF-28 for the telecom sources. The orthogonal linearly polarized fundamental modes of the fibers are determined by solving the propagation mode problem defined by the fiber geometry, where the fiber is considered to be infinitely extended (invariant) in the propagation direction. Subsequently, the fraction of the light field energy that is coupled into the twofold degenerate fundamental mode of the fiber is calculated. For this purpose, a mode overlap integral of the field scattered into the fiber and the fundamental mode is calculated [24].

Fig. 5b shows exemplarily the field distribution for the entire fiber-coupled micropillar emitting at 930 nm. On the left side, high field intensities can be found in the area of the micropillar and its radiation cone. The emitted light diverges and is focused on the fiber facet by the aspheric lens, which is located at an FEM-optimized distance of 38.5 µm from the planarized micropillar-air interface. It can be clearly seen that the light is guided in the fundamental modes of the fiber. In Fig. 5c, cross-sections of the field distributions of the propagating field and the fiber modes at the fiber facet level are shown for all structures emitting at 930 nm. The field distributions are approximately Gaussian shaped in each case. This is also reflected in the high coupling efficiency - defined as the ratio of the power guided into the fiber to the power incident on the fiber facet - of 89%, 92%, 90% and 95% for micromesa, microlens, CBG, and micropillar, respectively. The micropillar and micromesa feature emission with highest rotation symmetric Gaussian profile, with better mode overlap for the micropillar, resulting in the highest coupling efficiency for this structure.

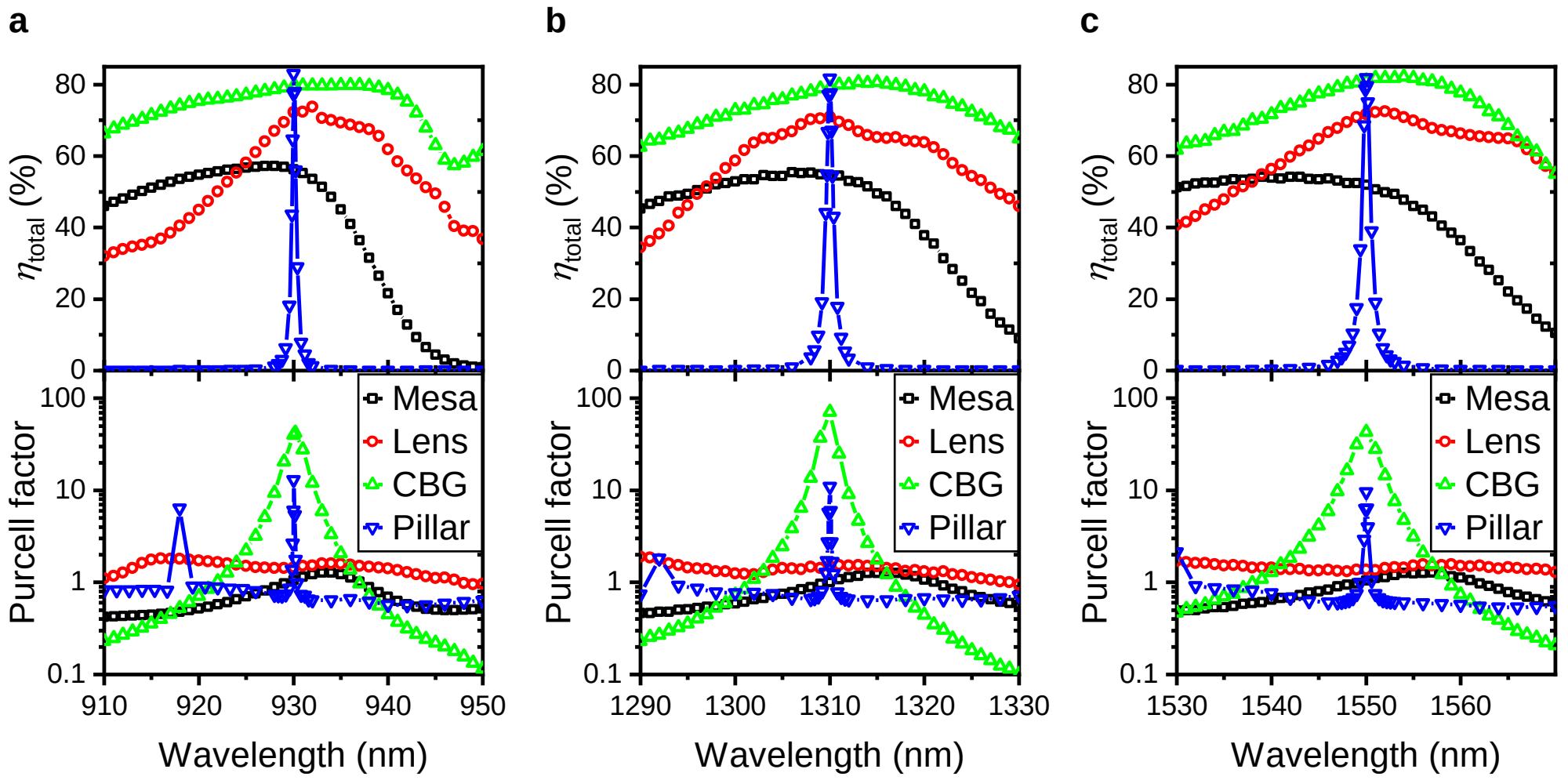

Fig. 6. FEM simulations of the full systems (semiconductor structure - coupling system - single mode fiber) with overall efficiencies (top) and Purcell factors (bottom) around 930 nm (a), 1310 nm (b) and 1550 nm (c).

To calculate the total efficiency $\eta_{total}$ of the entire fiber-coupled quantum-light-source system, the power guided into the fundamental fiber mode is set in relation to the total radiated power by the QD. The overall system efficiencies are shown for all systems and wavelengths studied in Fig. 6. These includes the $\eta_{ext}$ of the microstructure, reflections at the polymer lens (~4.5%) and fiber interface (<0.1%) and a possible influence of backscattered light on the cavity properties. It is again shown that the same structures at different wavelengths exhibit comparable performance. Thus, $\eta_{total}$ for the micromesa is on average 55%, for the microlens 72%, for the CBG structure 81%, and for the micropillar 83%. As previously mentioned, for selected quantum applications, a broadband enhancement of emission is desired, such as for the generation of polarization-entangled photon pairs from a biexciton-exciton cascade of a QD [39, 40]. In addition to $\eta_{total}$, Fig. 6 also shows $F_P$ for wavelength windows of 40 nm. Here it can be seen that the properties of the systems discussed in section 2.3 are largely preserved in the fiber-coupled case. Overall, the highest $\eta_{total}$ are achieved for the micropillars, with the CBG system being only 2 percentage points behind but showing significantly broader bandwidth and higher Purcell enhancement. The more elliptical mode profile and associated lower coupling efficiency is almost compensated by an increased $\eta_{ext}$. The fiber-coupled microlens is surprisingly efficient. However, it must be taken into account that such high-performance lenses have not yet been realized experimentally. In principle, however, it is also possible to achieve overall efficiencies of up to 56% with a simple mesa using an appropriate coupling geometry.

## 4. Conclusion

In summary we performed a comprehensive numerical study on the single-mode fiber-coupling of four different QD-based quantum-light source concepts. We focused on maximizing the overall photon-in-fiber-coupling efficiency by maximizing the photon extraction efficiency of the QD-sources and the fiber-coupling efficiency by including an intermediate lens to ensure mode matching between the source and the fiber core. The considered device concepts relying on a micromesa, a microlens, a micropillar, and a CBG as photon extraction strategies were modeled and optimized in detail by FEM simulations. For each device concept a specific aspherical lens was designed, which couples the far-field emission of the source into the guided modes of a single mode fiber. To foster the practical implementations of the fiber-coupling solutions the aspherical lenses are designed in such a way that they can be written onto the fiber facet using existing 3D direct-laser-writing technology [32, 23]. The printing of fiber holders and their accurate positioning relative to the structure has also been successfully demonstrated in the past [54, 23], so that the proposed overall systems can be implemented with existing working routines. Self-contained FEM simulations of high accuracy were used to evaluate the overall performance. This results in overall efficiencies between 55% for the mesa, 72% for the microlens, 81% for the CBG, and 83% for the micropillar. A further increase in efficiency of up to 4.5% is expected by using an appropriate anti-reflection coating for the aspherical lenses. In addition to overall efficiencies, also other device parameters should be taken into account when choosing a source: Micromesas and microlenses are very broadband and are comparatively straightforward to fabricate using in-situ lithography with high accuracy [53]. In contrast, CBG structures and the micropillar cavities are more useful for the generation of indistinguishable photons because of Purcell-reduced lifetime of the QD states [35, 36, 37, 39, 40]. Moreover, while the fiber-coupled micropillar has the highest overall efficiency, the CBG structure is ideal for generating polarization-entangled photon pairs due to the broadband cavity mode [39, 40].

The comparable and exceptional overall efficiencies for three wavelength ranges (NIR, telecom O- and C-band) and two fiber types underline the universal applicability of our coupling scheme. Moreover, these values exceed those achieved so far in experiments as well as in simulations in which selected sources were specifically adapted to a certain fiber type, which cannot be freely selected [24, 26, 25]. The simulations thus confirm that self-contained simulations of the whole system including the QD source, a mode-matching lens and the fiber are necessary to maximize the overall photon-in-fiber-coupling efficiency. The studied on-chip fiber-coupling of semiconductor quantum light sources is crucial for the development of practical quantum emitters and will help to fulfill the promise of highly efficient fiber-coupled sources, as currently urged by the quantum network community.

**Funding**

This work was supported by the German Federal Ministry of Education and Research (BMBF) through the projects Q.Link.X, QR.X. and PRINTOPTICS.

We acknowledge financial support by the 20FUN05 SEQUME project. This project (20FUN05 SEQUME) has received funding from the EMPIR programme co-financed by the Participating States and from the European Union's Horizon 2020 research and innovation programme.

We further acknowledge funding by the Baden-Württemberg (BW)-Stiftung through the project Opterial and from the European Research Council (ERC) through the project PoC Grant 3DPrintedOptics.

Funded by the Deutsche Forschungsgemeinschaft (DFG, German Research Foundation) under Germany´s Excellence Strategy – The Berlin Mathematics Research Center MATH+ (EXC-2046/1, project ID: 390685689) and by the DFG graduate school GRK 2642.**Acknowledgments** We thank for the possibility of using the computing resources of the HPC Cluster of the Faculty II of TU Berlin.

**Disclosures** The authors declare no conflicts of interest.

**Data availability statement** Data underlying the results presented in this paper are not publicly available at this time but may be obtained from the authors upon reasonable request.

# Numerical optimization of single-mode fiber-coupled single-photon sources based on semiconductor quantum dots: supplemental document

## Refractive indices for cryogenic temperatures

For the fibers, it was reasonably assumed that the silicon fiber cladding is undoped and the fiber core is doped. Based on the known refractive index of the cladding, the calculation of the refractive index of the core can be determined using $\mathrm{NA} = \sqrt{n_{\mathrm{core}}^2 - n_{\mathrm{cladding}}^2}$. The NA of the 780-HP fiber is assumed to be NA=0.1027 at 930 nm (given by the supplier on request). For the SMF-28, the NA is given as 0.14 in the data sheet for 1310 nm as well as for 1550 nm. Table S1 summarizes the values used.

**Table S1 Refractive indices for fiber core and cladding for the three wavelengths examined.**

| Wavelength (nm) | $n_{\mathrm{core}}$ @ 4 K | $n_{\mathrm{cladding}}$ @ 4 K [1] |
|---|---|---|
| 930 | 1.454 | 1.450 |
| 1310 | 1.453 | 1.446 |
| 1550 | 1.450 | 1.443 |

Tables S2-S4 show the used refractive indices of GaAs, AlAs and InGaAs, respectively, where the value at 4 K was always estimated from the value at 300 K based on the references given. The refractive index for the IPS-S polymer used was 1.503, 1.499 and 1.496 for 930 nm, 1330 nm, and 1550 nm, respectively [2].

**Table S2 Refractive indices of GaAs.**

| Wavelength (nm) | $n$ @ 300 K [3] | $n$ @ 4 K [4] |
|---|---|---|
| 930 | 3.5125 | 3.424 |
| 1310 | 3.4036 | 3.315 |
| 1550 | 3.3779 | 3.289 |

**Table S3 Refractive indices of AlAs.**

| Wavelength (nm) | $n$ @ 300 K [5] | $n$ @ 4 K [6] |
|---|---|---|
| 930 | 2.963 | 2.915 |
| 1310 | 2.908 | 2.860 |
| 1550 | 2.892 | 2.845 |

**Table S4 Refractive indices of InGaAs for different In fractions at 1550 nm.**

| Indium fraction | $n$ @ 300 K [7] | $n$ @ 4 K [8] |
|---|---|---|
| 36% | 3.49 | 3.40 |
| 29% | 3.46 | 3.37 |
| 10% | 3.39 | 3.30 |

### Telecom C-band structures

The C-band structures follow the design of a metamorphic buffer layer recently published by Sittig et al [9]. For the thickness of the jump layer, we have assumed a thickness of 30 nm with an indium content of 10%. Instead of the subsequent layer with increasing indium content, we assumed a fixed indium content of 36%, and the layer was 130 nm thick. A gradient in refractive index is implementable in JCMsuite but it is expected to have a negligible effect on the FEM results, so it was omitted. Above it is followed by the inverse sequence and a cap layer, respectively, with an indium content of 29.3%, in which the dipole is located.

## Radiation profiles

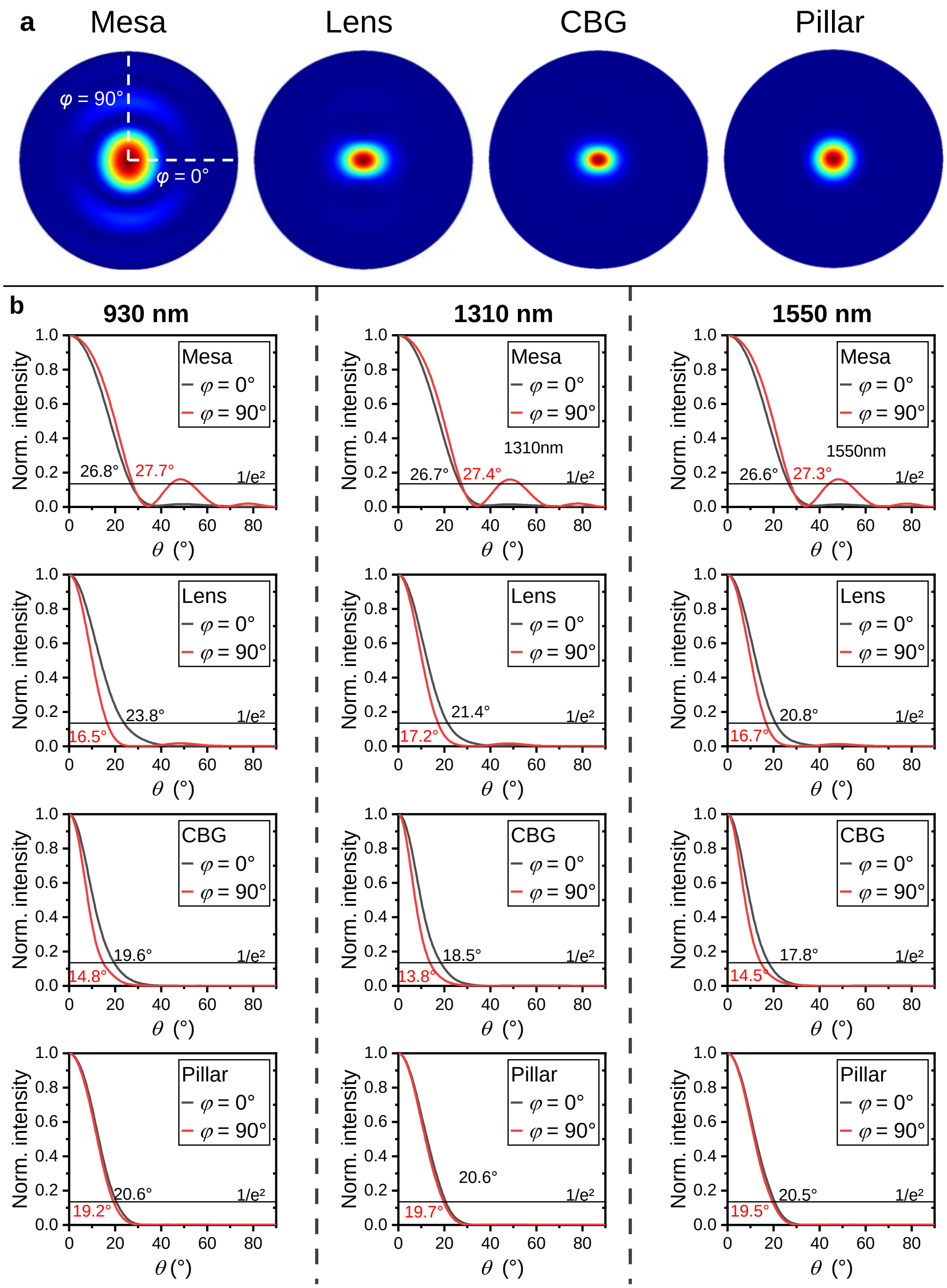

Figure S1 (a) Exemplary far-field distributions of the investigated structures emitting at 930 nm. The intensity distributions are shown logarithmically and cover the whole upper hemisphere. (b) Linescans along φ =0° and φ = 90° as a function of polar angle θ for all calculated far-field distributions.

## Parameters of the single-lens systems

The aspherical fiber lenses are defined by the following function:

$$z(r) = \frac{Cr^2}{1+\sqrt{1-(1+\kappa)C^2r^2}} + (a_2)r^2 + (a_4)r^4.$$

Here $z(r)$ describes the sag of the lens surface parallel to the optical axis as a function of the radial distance $r$ from the optical axis. $a_2$ and $a_4$ are the coefficients of the correction polynomial. $C$ describes the vertex curvature of the paraxial lens surface, while $\kappa$ describes the conical constant measured at the vertex. Below, Tables S5-S7 summarize the parameters of the one-lens systems for 930 nm, 1310 nm, and 1550 nm, respectively. The l*ens length* refers to the total height of the polymer imprint on the fiber lens, including the aspherically curved surface. *Distance to source* denotes the distance of the apex of the fiber polymer lens from the GaAs-air structural interface.

| Structure (930 nm) | Lens length (mm) | Distance to source* (mm) | Vertex radius (mm) | Conic constant | $a_2$ | $a_4$ |
|---|---|---|---|---|---|---|
| Mesa | 0.248 | 0.019 | 0.012 | -3.063 | 10.402 | 212.859 |
| Lens | 0.254 | 0.031 | 0.023 | -3.001 | 11.535 | -2715.720 |
| CBG | 0.240 | 0.037 | 0.027 | -1.302 | 9.766 | -3571.728 |
| Pillar | 0.259 | 0.039 | 0.026 | -2.746 | 9.381 | -433.597 |

Table S5 Parameters of the single-lens system for the structures emitting at 930 nm. The parameters were obtained by the procedure described in 3.1. *An exception is the distance to the source, which was optimized by means of a FEM parameter scan.

| Structure (1310 nm) | Lens length (mm) | Distance to source* (mm) | Vertex radius (mm) | Conic constant | $a_2$ | $a_4$ |
|---|---|---|---|---|---|---|
| Mesa | 0.205 | 0.020 | 0.013 | -3.324 | 12.444 | -10.869 |
| Lens | 0.214 | 0.037 | 0.023 | -3.441 | 9.157 | 1643.424 |
| CBG | 0.226 | 0.050 | 0.027 | -3.194 | 6.291 | 3555.523 |
| Pillar | 0.238 | 0.047 | 0.026 | -3.913 | 7.728 | 2355.705 |

Table S6 Parameters of the single-lens system for the structures emitting at 1310 nm. The parameters were obtained by the procedure described in 3.1. *An exception is the distance to the source, which was optimized by means of FEM parameter scans.

| Structure (1550 nm) | Lens length (mm) | Distance to source* (mm) | Vertex radius (mm) | Conic constant | $a_2$ | $a_4$ |
|---|---|---|---|---|---|---|
| Mesa | 0.213 | 0.052 | 0.013 | -2.912 | 11.271 | -120.122 |
| Lens | 0.214 | 0.036 | 0.023 | -3.435 | 8.456 | 2681.434 |
| CBG | 0.209 | 0.049 | 0.027 | -3.143 | 6.591 | 4383.544 |
| Pillar | 0.248 | 0.048 | 0.026 | -4.293 | 7.413 | 2448.459 |

Table S7 Parameters of the single-lens system for the structures emitting at 1550 nm. The parameters were obtained by the procedure described in 3.1. *An exception is the distance to the source, which was optimized by FEM parameter scans.

### Stability of the lens systems in terms of a vertical offset

While the photonic semiconductor structures and the aspherical lenses on the fibers can be evaluated independently after fabrication, this is hardly the case for the final distance between fiber lens and structure. The fiber is secured against lateral displacement by the fiber chuck in the given design [10], but the fulfillment of the desired distance from the fiber lens to the source is subject to a small uncertainty. Although the distance is determined by the height of the inner part of the chuck and the fiber is inserted until it reaches the rest point, the final fixture is provided by a drop of adhesive. It cannot be ruled out that when the adhesive cures, there will be a minimal change in distance. Therefore, the distance of the fiber lens from the structure was explicitly investigated. This investigation also serves to optimize the distance between the structure and the fiber lens using FEM parameter scans. Fig. S2 shows the relative decrease of the total efficiency in dB for deviations from the optimal distance for all investigated structures. The structures at 930 nm are more susceptible to such deviations, which can be explained by generally smaller distances between fiber lens and structure (see tables S5-S7). Furthermore, it is noticeable that the mesa structures are particularly sensitive to distance changes, since lenses of stronger refraction are used here, due to the radiation in larger solid angles. Essentially, the figure shows that the lens systems are more robust against a vertical displacement.

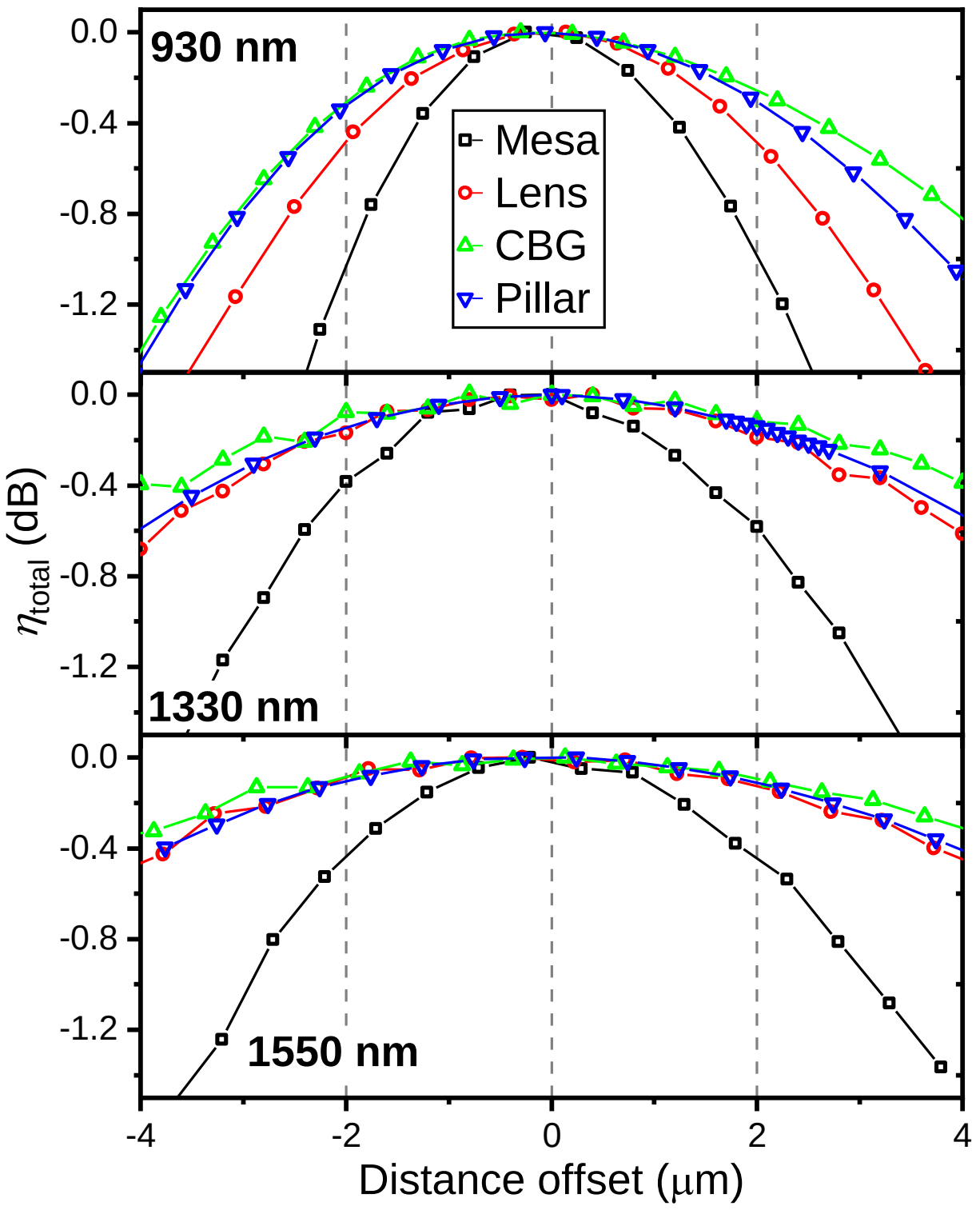

Figure S2 Overall efficiency of all studied structures as a function of the distance of the apex of the fiber lens from the GaAs surface